\documentclass[dvips]{acta}
\usepackage{supertabular,lscape,epsfig}
\usepackage{amssymb}
\usepackage{amsmath}
\DeclareSymbolFont{ppa}{OT1}{ppl}{m}{it}
\DeclareMathSymbol{\vv}{\mathalpha}{ppa}{'166}

\newfont{\hb}{rphvb at 10pt}
\newfont{\hbo}{rphvbo at 10pt}
\newfont{\bitt}{rptmbi at 12pt}
\newfont{\bits}{rptmbi at 11pt}

\SetPages{21}{36}

\SetVol{63}{2013}

\begin{document}

\newcommand{\TabApp}[2]{\begin{center}\parbox[t]{#1}{\centerline{
  {\bf Appendix}}
  \vskip2mm
  \centerline{\small {\spaceskip 2pt plus 1pt minus 1pt T a b l e}
  \refstepcounter{table}\thetable}
  \vskip2mm
  \centerline{\footnotesize #2}}
  \vskip3mm
\end{center}}

\newcommand{\TabCapp}[2]{\begin{center}\parbox[t]{#1}{\centerline{
  \small {\spaceskip 2pt plus 1pt minus 1pt T a b l e}
  \refstepcounter{table}\thetable}
  \vskip2mm
  \centerline{\footnotesize #2}}
  \vskip3mm
\end{center}}

\newcommand{\TTabCap}[3]{\begin{center}\parbox[t]{#1}{\centerline{
  \small {\spaceskip 2pt plus 1pt minus 1pt T a b l e}
  \refstepcounter{table}\thetable}
  \vskip2mm
  \centerline{\footnotesize #2}
  \centerline{\footnotesize #3}}
  \vskip1mm
\end{center}}

\newcommand{\MakeTableApp}[4]{\begin{table}[p]\TabApp{#2}{#3}
  \begin{center} \TableFont \begin{tabular}{#1} #4 
  \end{tabular}\end{center}\end{table}}

\newcommand{\MakeTableSepp}[4]{\begin{table}[p]\TabCapp{#2}{#3}
  \begin{center} \TableFont \begin{tabular}{#1} #4 
  \end{tabular}\end{center}\end{table}}

\newcommand{\MakeTableee}[4]{\begin{table}[htb]\TabCapp{#2}{#3}
  \begin{center} \TableFont \begin{tabular}{#1} #4
  \end{tabular}\end{center}\end{table}}

\newcommand{\MakeTablee}[5]{\begin{table}[htb]\TTabCap{#2}{#3}{#4}
  \begin{center} \TableFont \begin{tabular}{#1} #5 
  \end{tabular}\end{center}\end{table}}

\newfont{\bb}{ptmbi8t at 12pt}
\newfont{\bbb}{cmbxti10}
\newfont{\bbbb}{cmbxti10 at 9pt}
\newcommand{\uprule}{\rule{0pt}{2.5ex}}
\newcommand{\douprule}{\rule[-2ex]{0pt}{4.5ex}}
\newcommand{\dorule}{\rule[-2ex]{0pt}{2ex}}
\def\thefootnote{\fnsymbol{footnote}}
\begin{Titlepage}
\Title{The Optical Gravitational Lensing Experiment.\\
The OGLE-III Catalog of Variable Stars.\\
XV.~Long-Period Variables in the Galactic Bulge\footnote{Based on
observations obtained with the 1.3-m Warsaw telescope at the Las Campanas
Observatory of the Carnegie Institution for Science.}}
\vspace*{5pt}
\Author{I.~~S~o~s~z~y~ñ~s~k~i$^1$,~~
A.~~U~d~a~l~s~k~i$^1$,~~
M.\,K.~~S~z~y~m~a~ñ~s~k~i$^1$,~~
M.~~K~u~b~i~a~k$^1$,\\
G.~~P~i~e~t~r~z~y~ñ~s~k~i$^{1,2}$,~~
£.~~W~y~r~z~y~k~o~w~s~k~i$^{1,3}$,~~
K.~~U~l~a~c~z~y~k$^1$,~~
R.~~P~o~l~e~s~k~i$^{4,1}$,~~\\
S.~~K~o~z~³~o~w~s~k~i$^1$,~~
P.~~P~i~e~t~r~u~k~o~w~i~c~z$^1$~~
and~~J.~~S~k~o~w~r~o~n$^1$}
{$^1$Warsaw University Observatory, Al.~Ujazdowskie~4, 00-478~Warszawa, Poland\\
e-mail:
(soszynsk,udalski,msz,mk,pietrzyn,wyrzykow,kulaczyk,rpoleski,simkoz,pietruk,jskowron)\\
@astrouw.edu.pl\\
$^2$Universidad de Concepción, Departamento de Astronomia, Casilla 160--C, Concepción, Chile\\
$^3$Institute of Astronomy, University of Cambridge, Madingley Road, Cambridge CB3~0HA,~UK\\
$^4$Department of Astronomy, Ohio State University, 140 W.~18th Ave., Columbus, OH~43210, USA}
\Received{March 20, 2013}
\end{Titlepage}
\Abstract{The fifteenth part of the OGLE-III Catalog of Variable Stars
(OIII-CVS) contains 232\,406 long-period variables (LPVs) detected in the
OGLE-II and OGLE-III fields toward the Galactic bulge. The sample consists
of 6528 Mira stars, 33\,235 semiregular variables and 192\,643 OGLE
small amplitude red giants. The catalog data and data resources that are
being published include observational parameters of stars, finding charts,
and time-series {\it I}- and {\it V}-band photometry obtained between 1997
and 2009.

We discuss statistical features of the sample and compare it with
collections of LPVs in the Magellanic Clouds. The vast majority of red
giant stars in the Galactic bulge have an oxygen-rich chemistry. Mira
variables form a separate group in the period--amplitude diagram, which was
not noticed for oxygen-rich Miras in the Magellanic Clouds. We find a clear
deficit of long-secondary period stars toward the Galactic center compared
to the sample of Magellanic Clouds' LPVs.}{Stars: AGB and post-AGB --
Stars: late-type -- Stars: oscillations (including pulsations) -- Galaxy:
center}

\Section{Introduction}
Long-period variables (LPVs) constitute a very numerous class of variable
stars, because practically all the stars on the red giant branch (RGB) and
asymptotic giant branch (AGB) that are brighter than a given magnitude
limit show intrinsic variability. Generally, the brighter is a giant star,
the larger is the amplitude of the light variations, so this limiting
magnitude depends on the accuracy of our luminosity measurements. Studies
of LPVs allow us to address advanced questions regarding stellar
structures, evolution and pulsation theory. Variable red giants are also
promising distance indicators, giving the opportunity to study structures
of stellar environments where they appear.

The first significant sample of over 200 LPVs in the central region of the
Milky Way was found by Shapley and Swope (1934) and Swope (1935, 1938,
1940). The spatial distribution and mean magnitudes of this sample was used
to study the interstellar extinction toward the Galactic center. Baade
(1946) reported a discovery of about 100 LPVs (a detailed list of these
objects was given by Gaposchkin 1955) in the relatively low-extinction
region centered upon the globular cluster NGC~6522 and known today as
Baade's Window. Then, Ponsen (1957), Hoffleit (1957) and Lloyd Evans (1976)
increased the number of known LPVs in the Galactic bulge to about a half
thousand. Wood and Bessell (1983) studied 51 of LPVs in the Galactic center
and showed that variables with periods longer than 250~days are
significantly redder than LPVs in the solar vicinity as well as in the
Magellanic Clouds. This was interpreted as a result of higher metallicity
of stars in the Galactic center. Whitelock and Catchpole (1992) used Mira
stars to show that the bulge has a bar-like structure. A near-infrared
(near-IR) survey for LPVs in the inner bulge, within about 30~pc of the
Galactic center, was conducted by Glass \etal (2001), who reported the
discovery of 409 large-amplitude LPVs. Matsunaga \etal (2009) surveyed the
same region with deeper near-IR photometry and identified 1364 variables
which were used to estimate the extinction and distance to the Galactic
center.

Large-scale variability surveys, like MACHO and Optical Gravitational
Lensing Experiment (OGLE), greatly improved our knowledge of the properties
of pulsating red giants. Minniti \etal (1998) presented preliminary results
of the search for semiregular variables (SRVs) in the bulge using the
photometric data collected by the MACHO project. They reported the
discovery of about 2000 SRVs which followed two parallel sequences in the
period--color diagram. Alard \etal (2001) used MACHO photometry to study
the mass loss in about 300 AGB stars in the Galactic bulge. Glass and
Schultheis (2003) extended this sample to more than 1000 objects and showed
a series of four sequences in the $\log{P}$--$K$~magnitude plane -- similar
to those discovered by Wood \etal (1999) in the Large Magellanic Cloud
(LMC).

\hglue-5pt Photometric data collected during the second phase of the OGLE 
project (OGLE-II) were used by Wray, Eyer and Paczyñski (2004) to
identify and study over 15\,000 OGLE small amplitude red giants (OSARGs) in
the Galactic bulge. Wray \etal (2004), among others, showed that these
stars also trace the Galactic bar. Wo¼niak, McGowan and Vestrand (2004)
searched for rapid variability events in the Galactic bulge Mira stars
observed by the OGLE-II survey and reported no such behavior. Groenewegen
and Blommaert (2005) extracted from the OGLE-II databases 2691 Mira light
curves and used them to discuss the structure and distance of the Galactic
center. Matsunaga, Fukushi and Nakada (2005) compared the properties of
1968 Miras detected in the OGLE-II bulge fields with LPVs observed in the
Magellanic Clouds.

In this paper we present a catalog which dramatically increases the number
of known LPVs in the Galactic bulge -- from thousands to hundreds of
thousands. This catalog is a part of the OGLE-III Catalog of Variable Stars
(OIII-CVS). To date in this series we published, among others, the catalogs
of RR~Lyr stars (Soszyñski \etal 2011a) and Cepheids (Soszyñski \etal
2011c) in the Galactic bulge, and the catalogs of LPVs in the Large
(Soszyñski \etal 2009) and Small Magellanic Cloud (SMC, Soszyñski \etal
2011b). The long-term time-series photometry of all stars in the Catalog is
available from the OGLE Internet Archive.

\Section{Observational Data}
The OGLE-III survey toward the Galactic center lasted from 2001 to
2009. Observations were obtained with the 1.3~m Warsaw Telescope, located
at the Las Campanas Observatory in Chile. The observatory is operated by
the Carnegie Institution for Science. The telescope was equipped with the
mosaic camera consisting of eight $2048\times4096$ detectors, with a
combined field of view of $35\arcm\times35\zdot\arcm5$ yielding a scale of
approximately 0.26~arcsec/pixel. Details of the instrumentation setup can
be found in the paper by Udalski (2003). In this catalog we used also
observations collected between 1997 and 2000 in the course of the OGLE-II
project. By that time the Warsaw Telescope was equipped with the ``first
generation'' camera with a SITe $2048\times2048$ CCD detector working in
drift-scan mode (Udalski, Kubiak and Szymañski 1997).

In this study we analyzed exactly the same area as in the catalogs of
RR~Lyr stars and Cepheids in the Galactic bulge (Soszyñski \etal 2011ac)
with the total field of 68.7~square degrees. Most of the observations were
obtained with the Cousins {\it I} photometric band. From a few to several
dozen points were collected in the Johnson {\it V}-band. The number of {\it
I}-band observations and the time coverage significantly varies from field
to field -- from about 100 points collected over two years to more than
3000 observations obtained between 1997 and 2009 (OGLE-II +
OGLE-III). Additionally, for the purpose of the period determination, we
supplemented the light curves of some LPVs by adding the OGLE-IV
observations collected in 2011 and 2012, but this photometry is not
published in this catalog. Some stars were identified twice or more times,
because they were located in the overlapping regions between adjacent
fields. In such cases we selected for publication the light curve from only
one of these fields -- usually this one that consists of a larger number of
points.

The OGLE project routinely uses the technique of Different Image Analysis
(DIA, Alard and Lupton 1998, Wo¼niak 2000) to obtain accurate photometry of
monitored stars. The DIA reference frames for each field are constructed
from a stack of the $\approx10$ best images (with low seeing and low
background). Most of the stars brighter than about $I=12.5$~mag are
saturated. The saturation limit for the OGLE-II data is slightly brighter,
about $I=11.5$~mag. For about 1600 LPVs that are overexposed in the DIA
reference frames we provide the {\sc DoPhot} {\it I}-band photometry
(Schechter, Mateo and Saha 1993) and flag these stars in the Remarks of the
catalog. Note that only a small fraction of overexposed LPVs in the
Galactic bulge were included in our catalog. More information on the
photometric reduction pipeline, photometric calibrations and astrometric
transformations can be found in Udalski \etal (2008).

For the classification purposes we cross-matched our sample with the 2MASS
All-Sky Catalog of Point Sources (Cutri \etal 2003) obtaining near-IR $J$
and $K_s$ magnitudes. Following Dutra, Santiago and Bica (2002) we
assumed the relation between extinction and reddening for the 2MASS
magnitudes as
$$A_{K_s}=0.670E(J-K_s)$$
and we defined the reddening-free near-IR Wesenheit index as
$$W_{JK}=K_s-0.670(J-K_s).$$

\Section{Selection and Classification of Long-Period Variables}
LPVs in the Galactic bulge were identified with procedures similar to those
used in the Magellanic Clouds (Soszyñski \etal 2009, 2011b). First, we
calculated periodograms for each of the $3\times10^8$ {\it I}-band light
curves obtained by the OGLE-II and OGLE-III projects in the Galactic
bulge. We searched the frequency space from 0.0005 to 0.2~day$^{-1}$
(periods from 5 to 2000 days) with a resolution of $10^{-6}$~day$^{-1}$
using the {\sc Fnpeaks} code (Z. Ko³aczkowski, private communication). For
each star we found five periods, iteratively fitting and subtracting
third-order Fourier series from light curves folded with consecutive
periods. For each period we also determined and recorded an amplitude,
defined as a difference between maximum and minimum value of the
third-order Fourier series fitted to the residual light curve.

The selection and classification of LPVs was primarily based on the light
curve shapes. We visually inspected light curves of all stars brighter than
$I=13$~mag and for fainter objects we limited our sample to the light
curves with larger amplitudes of variability, larger scatter of the
observing points, or ratios of periods typical for OSARG variables. In the
selection procedure we also took into account the $(V-I)_0$ color index of
each star, dereddened with the reddening maps of Nataf \etal (2013).

Following the OGLE-III catalogs of LPVs in the Magellanic Clouds
(Soszyñski \etal 2009, 2011b) we divided our sample into three classes:
Miras, SRVs and OSARGs. Stars with peak-to-peak amplitudes of the {\it
I}-band light curves larger than 0.8~mag were classified as
Miras. Separation of SRVs and OSARG variables is more difficult, in
particular in the Galactic bulge, where period--luminosity sequences are
significantly broadened by the depth along the line of sight. We took into
account characteristic ratios of periods of OSARG stars, amplitudes, and
positions in the near-IR period--luminosity diagrams. However, we cannot
exclude the possibility that some of the LPVs in our catalog are
incorrectly classified, especially for objects with a small number of
observing points.
\vskip6pt
Similarly, our catalog may contain a limited number of objects other than
LPVs. For example, during our search we detected tens of thousands rotating
spotted stars. In Fig.~1 we show three example light curves of SRVs and
three light curves of spotted variables. Although phased light curves may
look similar in both classes of variable stars, because SRVs and spotted
variables exhibit similar periods and variable amplitudes, the difference
is visible in the unfolded light curves. LPVs generally have multi-periodic
light curves and their amplitudes change from cycle to cycle, while spotted
variables usually show only one (rotation) period and the variations of
amplitudes are much slower. Nevertheless, distinguishing between LPVs and
spotted variables may not be trivial for small-amplitude stars or for
objects with a small number of observations. Thus, we cannot exclude that
there is a number of spotted variables rather than pulsating red giants
among the smallest amplitude variables in our catalog. Also, our sample may
still contain a small number of young stellar objects, foreground red
dwarfs and background quasars.
\vskip6pt
In the catalogs of LPVs in the Magellanic Clouds (Soszyñski \etal 2009,
2011b) we divided our samples into oxygen-rich and carbon-rich stars
using their position in the optical Wesenheit index \vs near-IR
Wesenheit index diagram. In this catalog we cannot use the same method,
because of the considerable depth of the Galactic bulge along the line
of sight. However, in the Magellanic Clouds we noticed that both
spectroscopic types of AGB stars show different morphology of their
light curves (Soszyñski \etal 2011b). C-rich Miras and SRVs usually
exhibit irregular changes of their mean luminosity, while O-rich LPVs
show much more stable light curves with one or more pulsating modes. The
vast majority of the LPVs in the Galactic bulge show the latter
behavior, so we recognize them as O-rich giants. This confirms earlier
spectroscopic findings of an extremely low ratio of C-rich to O-rich
giants in the Galactic bulge (Blanco, McCarthy and Blanco 1984), which
is expected in high-metallicity environments. We found several SRVs and
Miras with light curves similar to C-rich giants in the Magellanic
Clouds and we flagged them as ``possible C-rich stars'' in the Remarks
of the catalog. Their status has to be confirmed spectroscopically. One
of the candidates from this list, namely OGLE-BLG-LPV-149402, has
recently been found as the first C-rich Mira in the Galactic bulge by
the spectroscopic survey of the symbiotic star candidates (Mi\-szalski,
Miko³ajewska and Udalski (2013).

\begin{figure}[t]
\centerline{\includegraphics[width=12cm, bb=55 100 555 750]{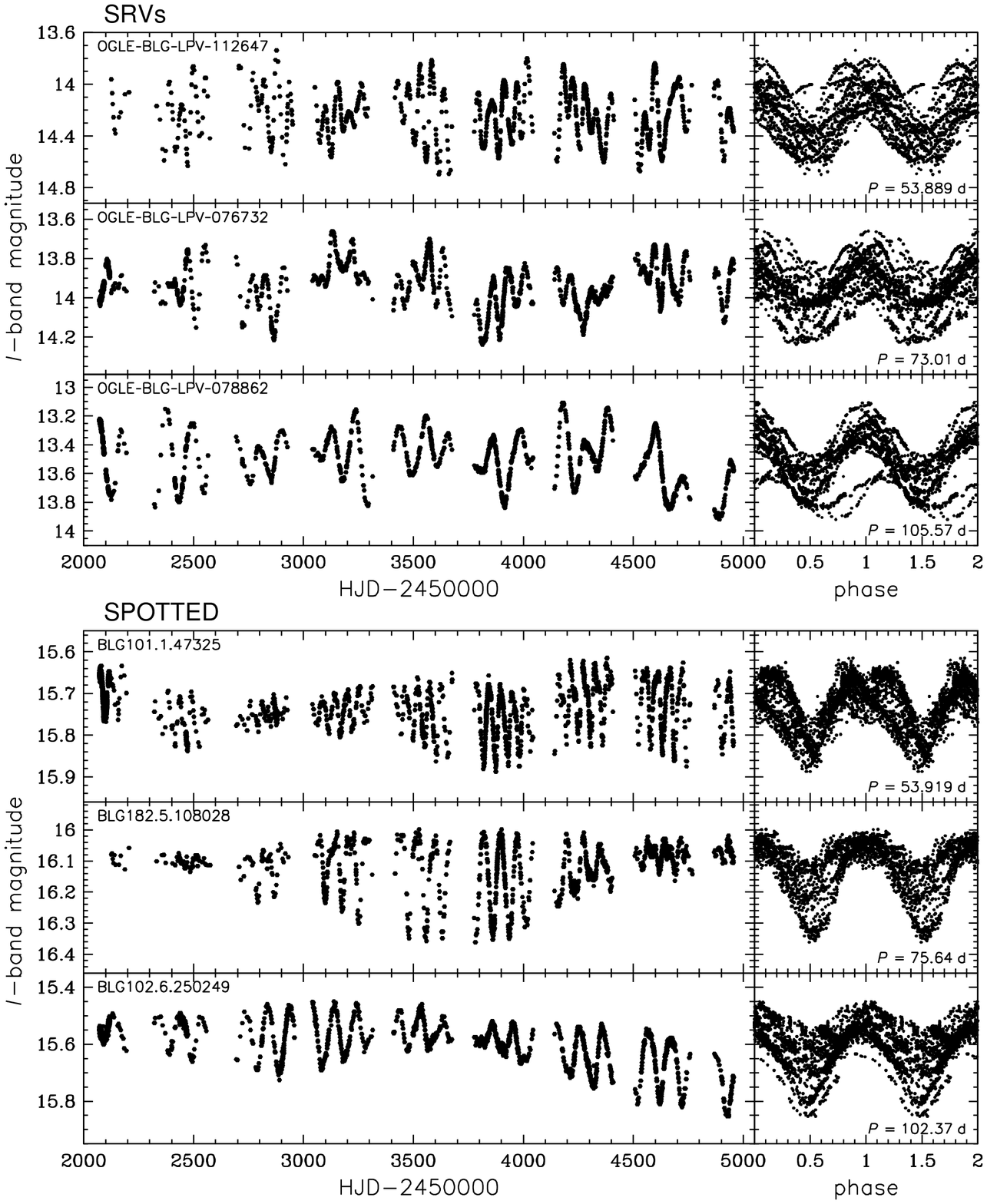}}
\FigCap{Example light curves of three SRVs ({\it upper panels}) and three
spotted stars ({\it lower panels}). {\it Left panels} show unfolded light
curves, and {\it right panels} show the same light curves folded with the
pulsation (SRVs) or rotation (spotted stars) periods. Note the difference
in amplitude variations visible in the unfolded light curves.}
\end{figure}
\begin{figure}[htb] 
\centerline{\includegraphics[width=12cm, bb=65 370 555 755]{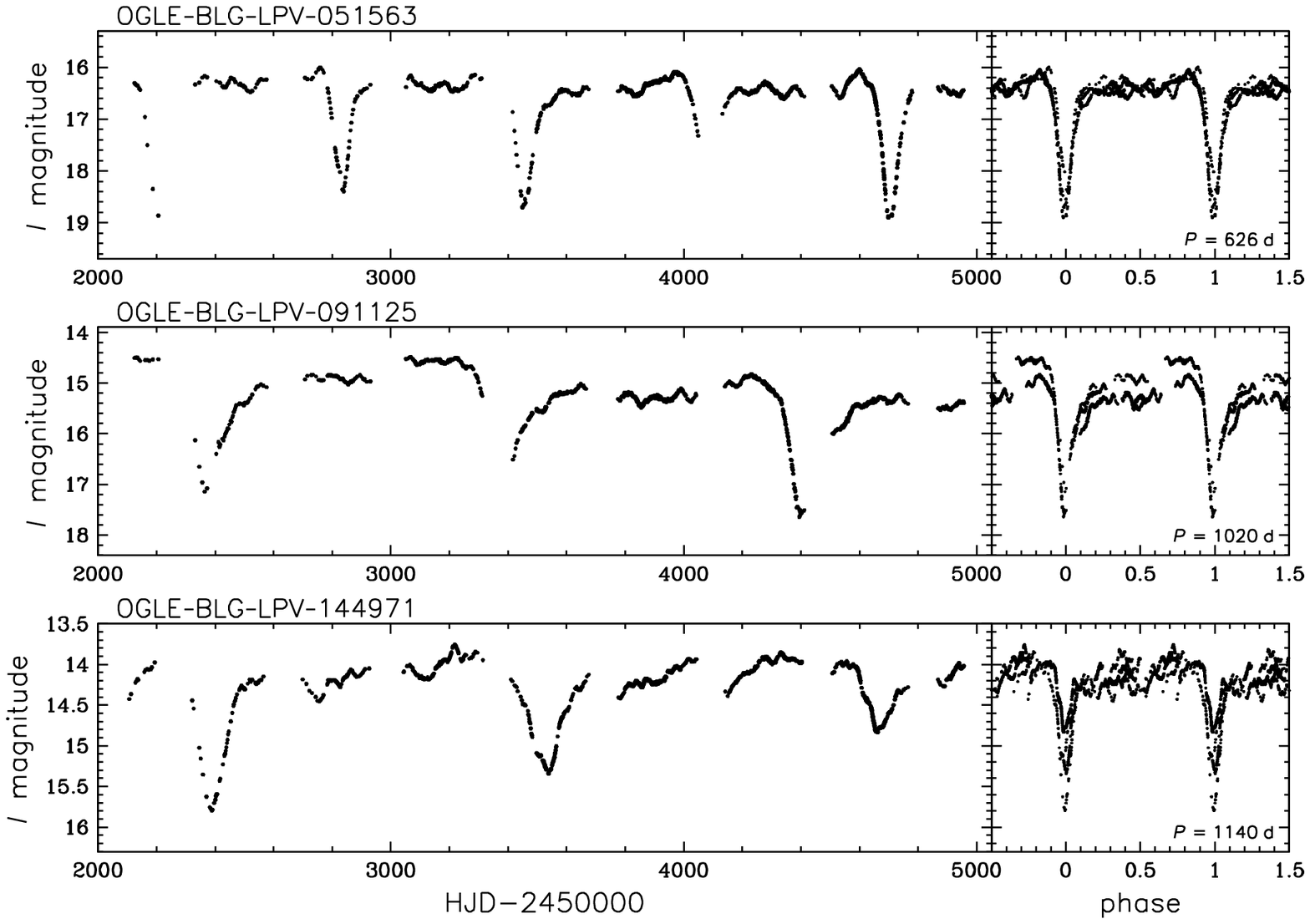}} 
\FigCap{Example light curves of LPVs with very deep, regular,
eclipsing-like minima in their light curves. {\it Left panels} show
unfolded light curves, while {\it right panels} show the same light
curves phased with the periods of the eclipsing-like minima.}
\end{figure} 
The catalog contains some red giants which exhibit eclipsing or ellipsoidal
modulation of their light curves, \ie which are members of close binary
systems. We left these objects on our list only when they simultaneously
showed pulsations as OSARGs or SRVs. The primary periods of these stars
given in the catalog usually correspond to half of the orbital periods,
because such periods were automatically found by our period-searching
code. About 800 distinct eclipsing or ellipsoidal red giants are flagged in
the Remarks. Besides close binary systems, our catalog contains several
SRVs and OSARGs that show eclipsing-like relatively narrow minima often of
a very large depth (>1~mag). Three example light curves of such objects are
shown in Fig.~2. It is worth noting that some of these objects may
belong to the symbiotic stars class. Miszalski \etal (2013) recently
presented a spectroscopically selected sample of symbiotic star
candidates containing many objects listed in our catalog.

During the search for LPVs in the Galactic bulge several gravitational
microlensing events with LPV Galactic bulge targets were serendipitously
found. This is not surprising as with the OGLE discovery rate of
microlensing phenomena of non-variable stars toward the Galactic center
(OGLE EWS System, Udalski 2003) one can expect significant sample of
microlensing events of variable sources as well. All microlensing events
found during our search are marked in the Remarks. Fig.~3 shows two
spectacular examples of microlensing phenomena that occurred on OSARG
variables.

\begin{figure}[htb] 
\centerline{\includegraphics[width=11.8cm, bb=30 240 570 755]{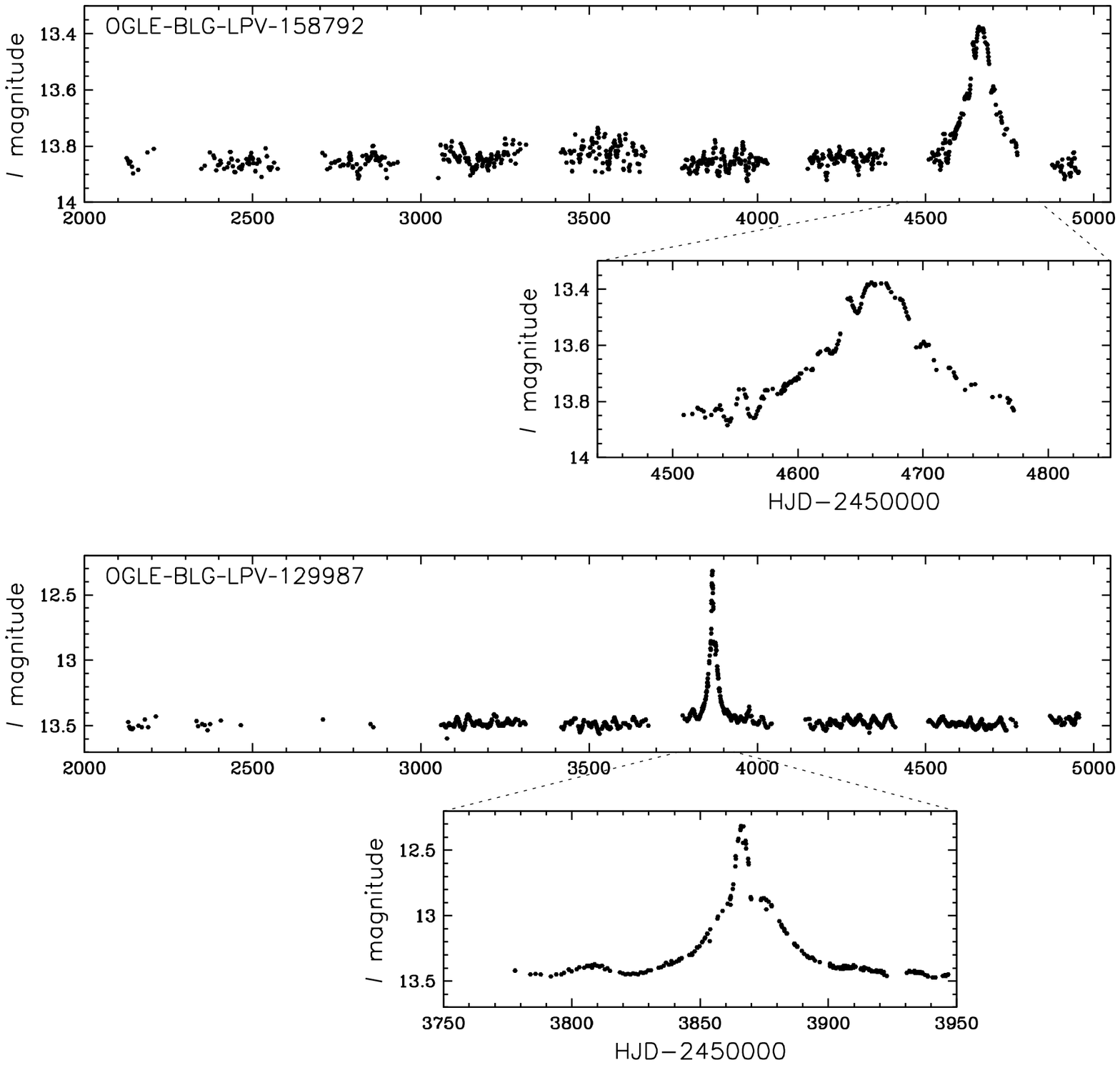}} 
\vskip5pt
\FigCap{Light curves of two gravitational microlensing phenomena
on OSARG variables: OGLE-BLG-LPV-158792 and OGLE-BLG-LPV-129987}
\end{figure} 

\Section{Catalog of Long-Period Variables in the Galactic Bulge}
The catalog of LPVs in the Galactic bulge contains 232\,406 objects, of
which 6528 have been classified as Miras, 33\,235 as SRVs and 192\,643 as
OSARGs. The proportion of different types of LPVs is similar to those in
the Magellanic Clouds, with somewhat larger fraction of Miras (2.8\% of the
total sample in the Galactic bulge \vs 1.8\% in the Magellanic Clouds), and slightly
larger fraction of SRVs (14\% \vs 12\%). These differences may reflect the
difficulties in classification of stars with the smallest amplitudes
(OSARGs) in the regions of high interstellar extinction or in the fields
with a relatively small number of observations (fields in the Magellanic
Clouds were monitored much more homogeneously).

The catalog and data on particular objects are accessible through the
anonymous FTP site or {\it via} the WWW interface:
\begin{center}
{\it ftp://ftp.astrouw.edu.pl/ogle/ogle3/OIII-CVS/blg/lpv/}\\
{\it http://ogle.astrouw.edu.pl/}\\
\end{center}

The FTP site is organized as follows. The lists of LPVs with their J2000
equatorial coordinates, classification, identifications in the OGLE-II and
OGLE-III databases and in the General Catalogue of Variable Stars (Samus
\etal 2011) are given in the {\sf ident.dat} file. The stars are arranged
in order of increasing right ascension and designated OGLE-BLG-LPV-NNNNNN,
where NNNNNN is a six-digit consecutive number.

\begin{figure}[p]
\hglue-2mm{\includegraphics[width=13.3cm]{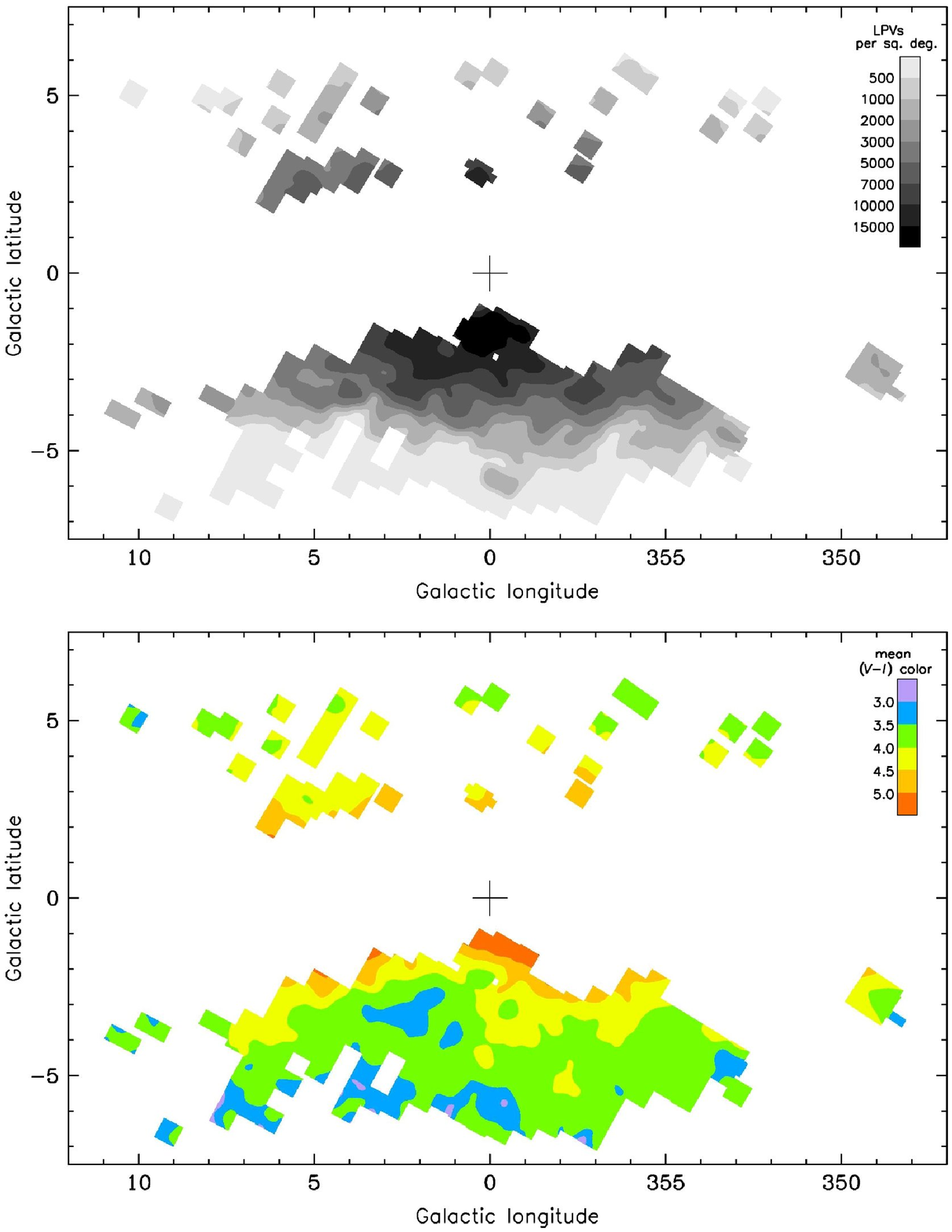}}
\vskip3mm
\FigCap{{\it Upper panel}: surface density map of LPVs in the Galactic
bulge. The map was obtained by blurring the spatial distribution of LPVs
with a Gaussian filter. {\it Lower panel}: spatial distribution of the mean
apparent $(V-I)$ colors of LPVs in the Galactic bulge.}
\end{figure}

The observational parameters of each star -- mean magnitudes in the {\it
I}- and {\it V}-bands, periods, and amplitudes -- are given in the files
{\sf Miras.dat}, {\sf SRVs.dat} and {\sf OSARGs.dat}, separately for each
type of LPVs. The files contain three periods per star found with the {\sc
Fnpeaks} code. Only for larger-amplitude variables and only the first
period was visually verified and corrected, if necessary. In most LPVs the
first given period corresponds to the pulsation period, or half of the orbital
period (for binary systems), or the long secondary period. The second and
the third periods provided in the catalog files were calculated fully
automatically and may be spurious in some cases. For the in-depth analysis
of the periodicity of LPVs from our catalog we recommend the reader to
perform an independent frequency search using multi-epoch OGLE photometry.

The time-series {\it I}- and {\it V}-band photometry is given in separate
files in the subdirectory {\sf phot/}. The subdirectory {\sf fcharts/}
contains finding charts for all objects. These are $60\arcs\times60\arcs$
subframes of the {\it I}-band DIA reference images, oriented with North up,
and East to the left. The file {\sf remarks.txt} contains additional
information about some objects.

The completeness of our catalog strongly depends on the brightness of
stars, their amplitude of light variability, interstellar extinction toward
a given star, the number and time span of observations of a given field,
etc. We checked the general efficiency of our variable star selection by
comparing objects located in the overlapping parts of adjacent
fields. Assuming that the minimum number of observing points must be larger
than 100, in total 10\,788 LPVs from our catalog were recorded in the OGLE
databases twice -- in the neighboring fields, so we had an opportunity to
independently detect 21\,576 counterparts. During the search we found
19\,285 of them which corresponds to the 89\% completeness of the total
sample. We carefully looked through the light curves that had been missed
during the search and we found that most of these objects had very small
amplitudes, usually at the detection limits of the OGLE photometry. The
same method applied only to SRVs and Miras gave much larger completeness --
above 98\%. However, one should keep in mind that these values concern only
those stars that can be detected by OGLE, with luminosities below the
saturation limit and with amplitudes above a few milimagnitudes.

Upper panel of Fig.~4 shows how the LPVs in our catalog are distributed in
the sky. This is a Gaussian-smoothed stellar density map. As can be seen,
the number of LPVs rises toward the Galactic center, with the exception of
the regions that are closest to the center, where the large extinction
prevents detection of LPVs with small amplitudes. In the lower panel of
Fig.~4 we present a map of the mean apparent $(V-I)$ colors of our red
giants. This map illustrates the spatial distribution of the interstellar
extinction toward the Galactic center.

\vspace*{-9pt}
\Section{Discussion}
\vspace*{-5pt}
The OGLE-III catalog of LPVs in the Galactic bulge is probably the largest
single collection of variable stars ever published. Together with the
already published parts of the OIII-CVS our catalog contains in total over
400\,000 variable stars in the Galactic bulge and Magellanic Clouds. This
huge sample of variable stars may be used for various studies of the
features and evolution of the stars themselves, as well as the structure and
history of the stellar environments in which they are observed.

Distribution of variable red giant stars in the period--luminosity (PL)
plane reveals a rather complex picture. Wood \etal (1999) discovered
five sequences in the PL diagram for LPVs and since that time the number
of known PL relationships obeyed by LPVs grows continuously (\cf
Soszyñski and Wood 2013). Red giants are often affected by significant
circumstellar extinction and the PL relations are the narrowest in the
near-IR band-passes, in particular in the reddening-independent
Wesenheit index $W_{JK}$. In Fig.~5 we plot the PL diagrams for LPVs in
the Galactic bulge using the 2MASS $K_s$ band and $W_{JK}$. Each star is
represented by only one (the primary) period. The well-known series of
PL sequences observed in the Magellanic Clouds is blurred by the larger
depth along the line of sight, however one can easily distinguish
sequence~C occupied by Miras and SRVs, sequence~C$'$ populated by SRVs and
sequence~D formed by still unexplained long secondary periods (LSPs).
The PL relations of OSARG variables overlap each other, but they may be
to some extent separated by means of the characteristic period ratios of
these multi-periodic oscillators. LPVs are promising distance
indicators, however all previous distance estimations to the Galactic
center from LPVs (Glass and Feast 1982, Glass \etal 1995, Groenewegen
and Blommaert 2005, Matsunaga \etal 2009) used solely Mira stars, while
SRVs and OSARGs also may be used for this purpose. Low-amplitude LPVs
are much more numerous than Miras and follow a series of well defined PL
relations. 

\begin{figure}[p] \centerline{\includegraphics[width=13cm]{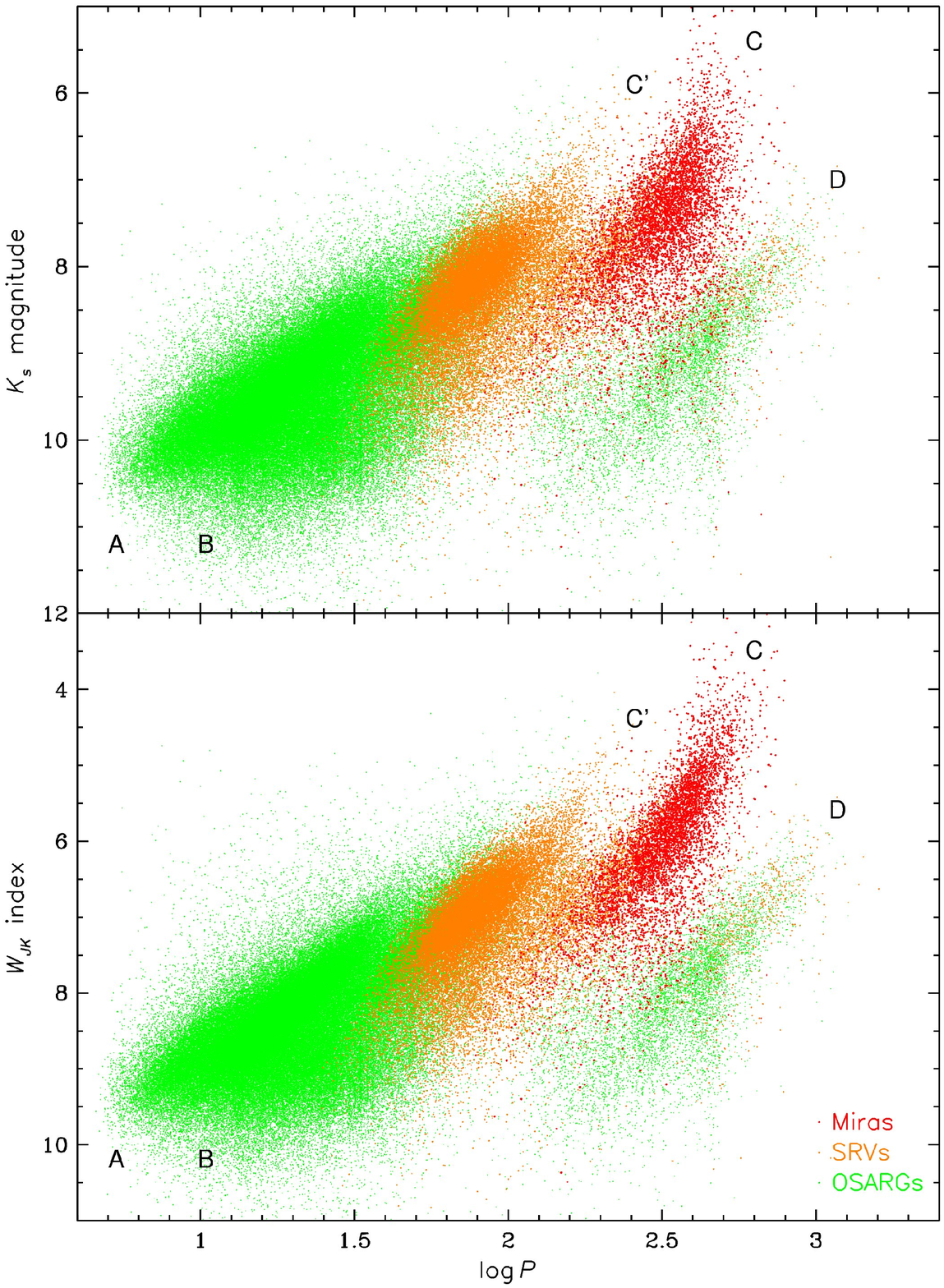}} 

\FigCap{Period--luminosity diagrams for LPVs in the Galactic bulge. {\it
Upper panel} shows $\log{P}{-}K_s$ diagram, while {\it bottom panel}
presents $\log{P}{-}W_{JK}$ diagram, where $W_{JK}$ is the
reddening-free Wesenheit index. Red points mark Mira stars, orange --
SRVs, and green -- OSARG variables. Each star is represented by one (the
primary) period.} 

\end{figure}

\begin{figure}[htb] 
\centerline{\includegraphics[width=10cm, bb=35 55 520 735]{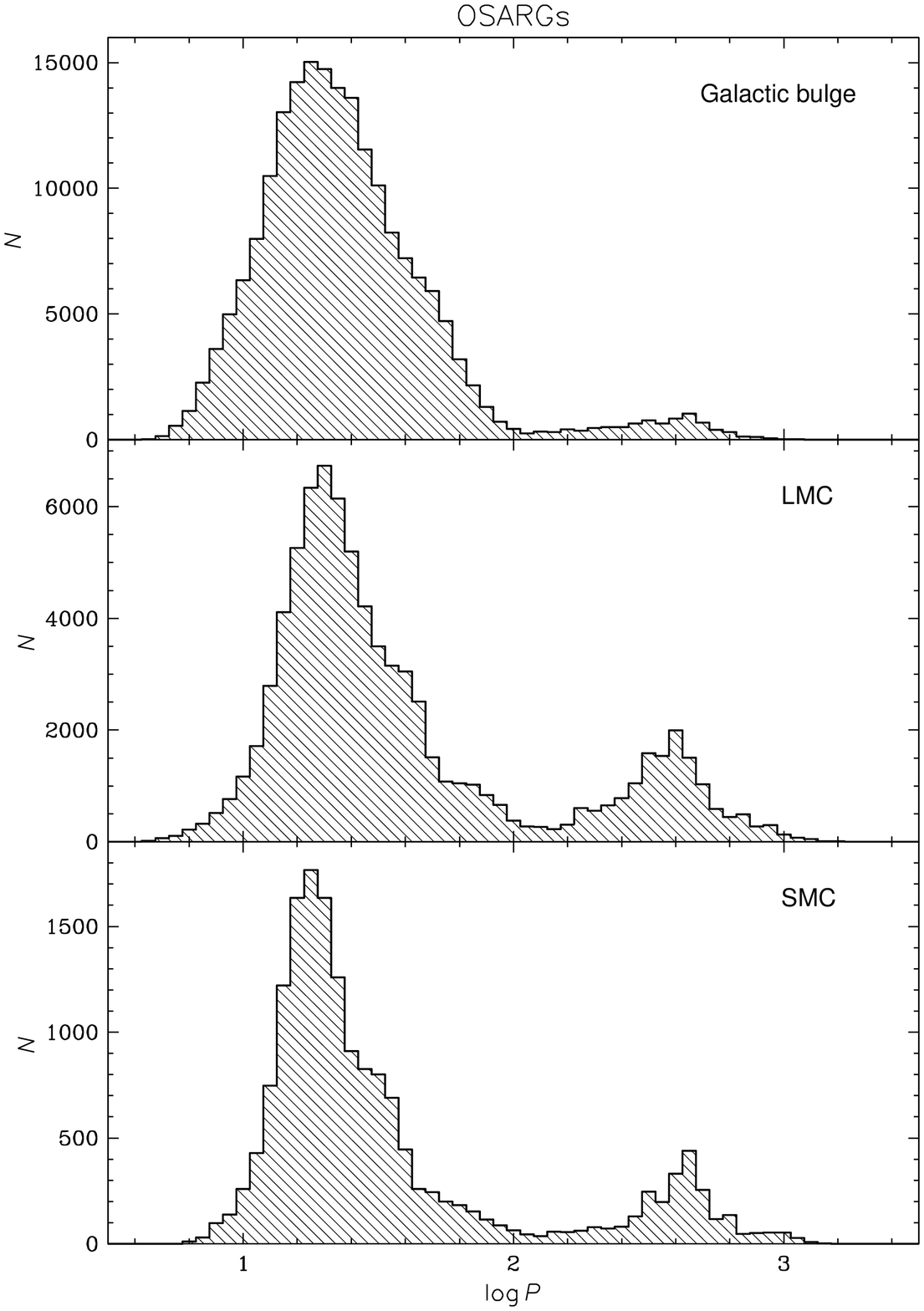}} 
\FigCap{Distribution of the primary periods of OSARG
variables in the Galactic bulge ({\it upper panel}), LMC ({\it middle
panel}) and SMC ({\it bottom panel}).} 
\end{figure} 
\begin{figure}[htb]
\centerline{\includegraphics[width=10cm, bb=35 55 520 735]{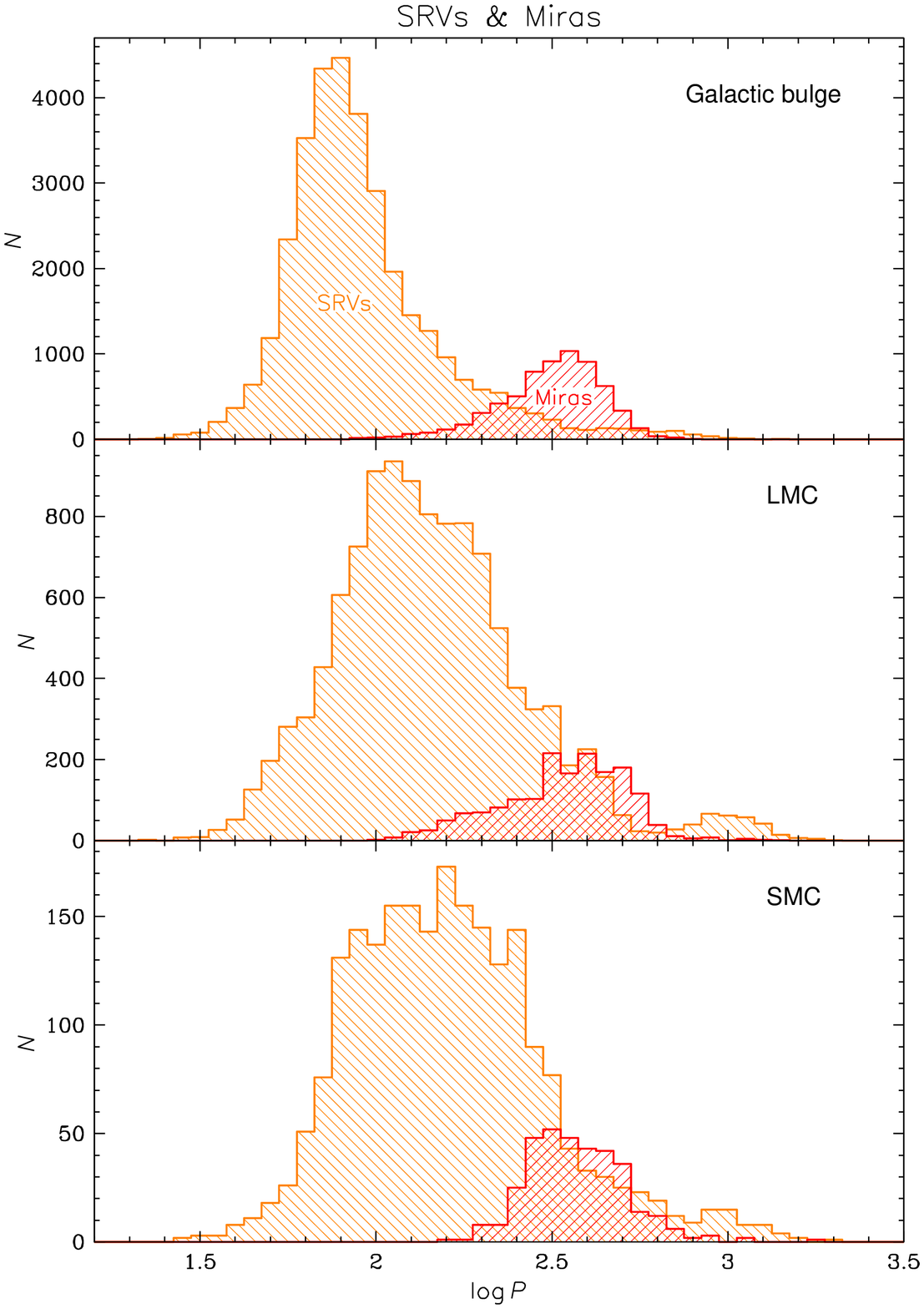}}
\FigCap{Distribution of the primary periods of SRVs (orange) and Miras
(red) in the Galactic bulge ({\it upper panel}), LMC ({\it middle
panel}) and SMC ({\it bottom panel}).} 
\end{figure}

Histograms showing the distribution of primary periods of OSARGs in the
Galactic bulge and in the Magellanic Clouds (Soszyñski \etal 2009, 2011b)
are plotted in Fig.~6. The striking difference between OSARGs in the
Galactic center and in the Magellanic Clouds appears for stars with LSPs
($2.2<\log{P}<3.1$). In the Galactic bulge, giants exhibiting this mysterious
phenomenon are much less frequent than in the Magellanic Clouds. We tested
if this could be an effect of shorter monitoring times of the subset of
fields toward the Galactic bulge. To do this we have constructed period
distribution for OSARG variables which had 5~years or longer light
curves. We noted virtually no difference between this distribution and the
full sample.

The period distribution of Mira stars (Fig.~7) has a similar shape in the
Galactic bulge and Magellanic Clouds. Note that these three environments
host completely different proportion of O-rich and C-rich Miras which is a
result of different metallicity, but this feature does not influence
significantly the periods of pulsations in the last stages of stellar
evolution on the AGB. On the other hand, the period distribution of SRVs
distinctly tends to be more concentrated with increasing metallicity. More
SRVs in the Galactic bulge have primary periods falling on sequence~C$'$
(first-overtone mode of pulsations), however one must remember that most of
them show multi-periodic behavior, and one of the secondary periods usually
falls on sequence~C (fundamental mode). It cannot be ruled out that
differences in primary period distributions of SRVs in the Galactic bulge
and Magellanic Clouds result from different time span of observation of fields in
these objects.
\begin{figure}[htb]
\includegraphics[width=12.93cm]{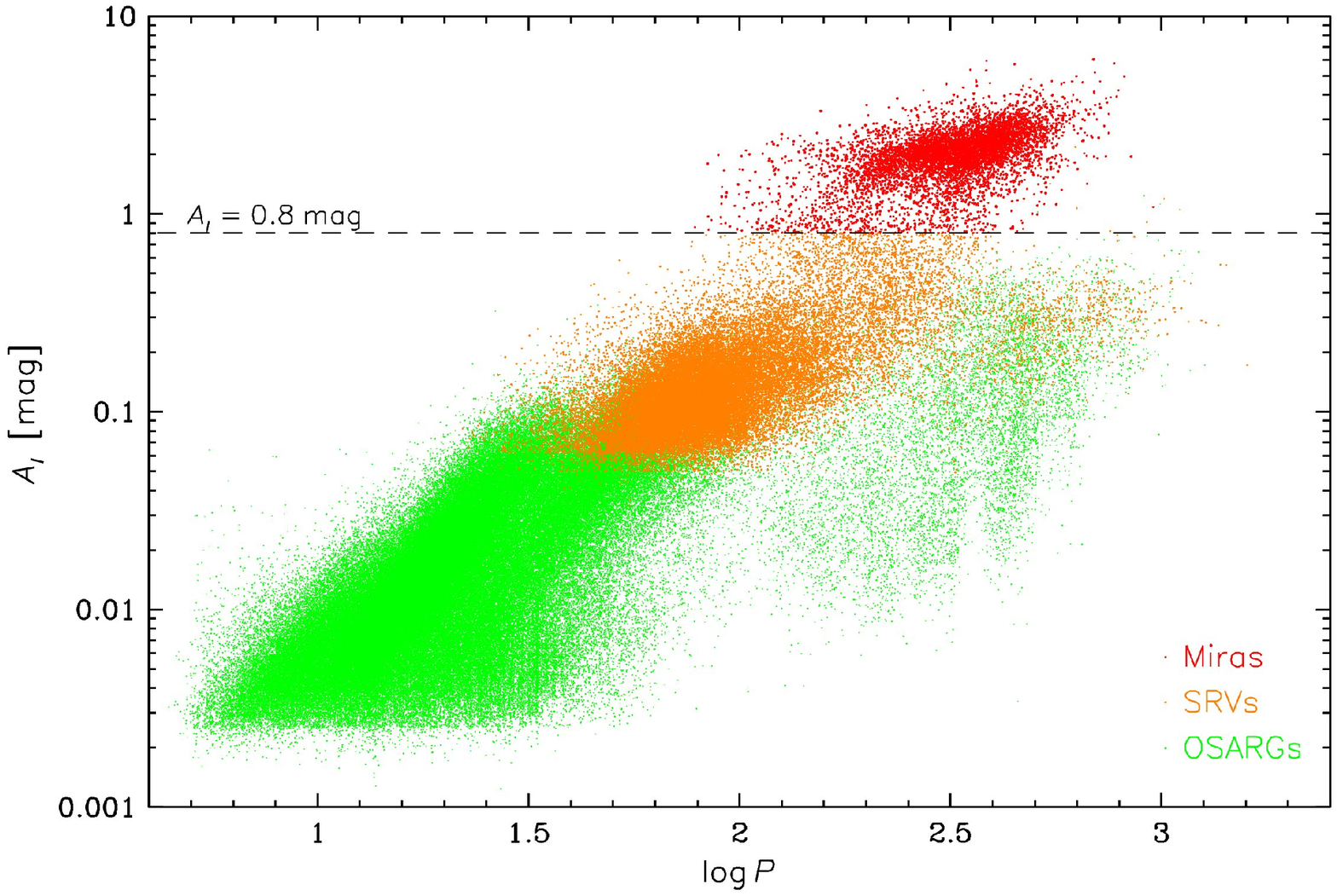}
\vskip6pt
\FigCap{Period--amplitude diagram for LPVs in the Galactic bulge. Red
points mark Mira stars, orange -- SRVs, and green -- OSARG variables. Each
star is represented by one (the primary) period.}
\end{figure}

In the period--amplitude diagram (Fig.~8) OSARG variables concentrate on
two sequences noticed by Minniti \etal (1998) and Wray \etal (2004). SRVs
partly overlap with OSARGs, while Mira stars clearly show a separate group
with {\it I}-band peak-to-peak amplitudes between 1.4~mag and 3~mag. We
noticed such a natural distinction between Miras and SRVs in the Magellanic
Clouds (Soszyñski \etal 2009, 2011b), but only for C-rich variables, while
in the Galactic bulge most of the SRVs and Miras are O-rich giants.

\vskip7pt
\Acknow{We thank Z.~Ko³aczkowski for providing software which enabled us
to prepare this study.

This work has been supported by the Polish Ministry of Science and Higher
Education through the program ``Ideas Plus'' award No. IdP2012 000162.

The research leading to these results has received funding from the
European Research Council under the European Community's Seventh Framework
Program\-me (FP7/2007-2013)/ERC grant agreement no.~246678.}

\end{document}